\documentclass[runningheads,fleqn]{cl2emult}

\usepackage{makeidx}  
\usepackage{graphicx} 
\usepackage{subeqnar} 
\usepackage{multicol} 
\usepackage{cropmark} 
\usepackage{phys}     
\makeindex            

\begin{document}
\title*{Introduction to the Propp-Wilson Method of Exact Sampling for 
the Ising Model}
\toctitle{Introduction to the Propp-Wilson Method of Exact Sampling 
\protect\newline for the Ising Model}
%
%
\titlerunning{Introduction to the Propp-Wilson}
%
\author{
M.~A.\ Novotny
}
\authorrunning{M.~A.\ Novotny}
%
%
\institute{
Supercomputer Computations Research Institute, 
Florida State University, Tallahassee, FL 32306-4130, USA
}

\maketitle              

\begin{abstract}
An introduction to the Propp-Wilson method of 
coupling-from-the-past 
for the Ising model is presented.  It enables one to obtain exact samples 
from the equilibrium spin distribution for ferromagnetic 
interactions.  Both uniform and random quenched magnetic fields are included.  
The time to couple from the past and its standard deviation are shown as 
functions of system size, temperature, and the field strength.  
The time should be an estimate of the ergodic time for the Ising system.  
\end{abstract}

The standard importance-sampling Monte Carlo algorithm 
\cite{refBH} for statics 
has a number of difficulties.  
One is its inability to ensure that 
the generated spin configurations are uncorrelated and are representative 
of the equilibrium distribution.  The standard 
way to overcome this difficulty is to throw away a large number of 
Monte Carlo steps between analyzed configurations.  Of course 
this decreases the number of configurations 
obtained in a given amount of computer time.  In addition, 
there can remain the question of whether the generated spin configurations 
are indeed drawn from the underlying equilibrium distribution.  
For example, consider a spin-glass model at low temperatures, 
where ergodicity has been broken for reasonable simulation times 
and the system might have many ground states and many more 
metastable states.  Then it is extremely difficult to decide 
whether the standard Monte Carlo method 
obtains spin configurations drawn from the underlying equilibrium spin 
distribution.  

In this short paper the square-lattice Ising model with 
nearest-neighbor ferromagnetic ($J$$>$$0$) interactions and periodic boundary 
conditions is considered.  The Hamiltonian is 
$
{\cal H} = - J \sum_{\langle i j \rangle} \sigma_i \sigma_j 
- \sum_i H_i \sigma_i ,
$
with $\sigma_i$$=$$\pm 1$, the first sum over nearest-neighbor spins, 
and $H_i$ is the local field.  
Two cases are considered, uniform field $H_i$$=$$H$, and quenched 
random field with $H_i$ chosen to be $\pm$$H$ with equal probability.  
We take $J$$=$$1$, 
which makes the exact critical temperature $T_{\rm c}$$\approx$$2.26\cdots$.  

One Monte Carlo step consists of two procedures.  First 
randomly choose a spin $i$ from the $L$$\times$$L$ lattice.  
The probability of 
$\sigma_i$ being set to $+1$ in the next time step is 
\begin{equation}
\label{EqProb}
{\rm Prob}(\sigma_i\!=\!1) = 
\left\{1+\exp\left[\left(-2 J \sum_{j={\rm nn}}\sigma_j 
-2 H_i\right) \Bigg / T\right]\right\}^{-1}  ,
\end{equation}
where the first sum is over the nearest-neighbors of spin $i$ 
and $T$ is the temperature.  
The second step is to use a uniformly distributed random 
number $r$, and set $\sigma_i$$=$$+1$ in the 
next time step if $r$$\le$${\rm Prob}(\sigma_i\!=\!1)$; otherwise 
set $\sigma_i$$=$$-1$.  
This is the single-spin flip heat-bath dynamic, which for the Ising 
model is the Glauber dynamic.  
I use the Ziff four-tap random number generator \cite{Ziff}.  
It is important to note that we do {\it not\/} consider the probability 
that a spin flips, but rather the probability of the spin being up, 
{\it i.e.\/} $+1$.  

Recently researchers in the statistics community have devised methods 
that can guarantee that generated spin configurations from a 
Monte Carlo simulation are drawn from the underlying equilibrium distribution.  
There the study goes under the name of 
Markov random fields \cite{refBk1}, which should {\it not\/} be confused 
with the random fields in the Hamiltonian.  
The method used here is the 1996 method of Propp and Wilson \cite{refPW}.  
It is referred to as `coupling-from-the-past' or the 
`Propp-Wilson method'.  
Here we bring the single-spin-flip version of the Propp-Wilson method 
to the attention of practitioners in the statistical mechanical community.  
We obtain a time for coupling from the past, which we 
call the Propp-Wilson time, $t_{\rm PW}$.  
We will show how $t_{\rm PW}$
changes with $T$, $L$, $H$, and the randomness in $H$.  
The time $t_{\rm PW}$ can be viewed as an upper bound 
to the ergodic time $\tau_{\rm e}$ of the simulation \cite{refBH}.  
In the spirit of a workshop, this study was only 
initiated about one month prior to the workshop.  Consequently, it 
should be viewed as an exploratory and preliminary work.  

The basic idea of the Propp-Wilson method is to not run the Markov chain 
forward in time (which all normal importance-sampling Monte Carlo 
procedures do), 
but rather to run the Markov chain from a time $-t$ in the past up to 
$t$$=$$0$.  Typically 
it is not known how to go from a spin configuration at 
one time to one at an earlier time.  However, given a spin configuration 
at time $-t$, it is easy to propagate it to time $t$$=$$0$ --- just 
use the algorithm for forward propagation.  What is the difference 
between going from $-t$ to $0$ or going from $0$ to $t$?  It lies only 
in our knowledge of 
the sequence of random numbers used.  At $t$$=$$0$ 
one has no idea which spin will be picked next or which random 
number the spin-up probability will be compared with.  However, at 
all times in the past, both the spin picked 
and the random number to compare with are known.  

Now consider starting two spin configurations at time $-t$, and propagating 
forward in time toward $t$$=$$0$.  The {\it same\/} random numbers 
are used for each spin configuration, since the same history of random 
numbers applies for times $t$$\le$$0$.  
At some time $-t_{\rm c}$ these two spin configurations might 
coalesce, i.e.\  they both become the same 
spin configuration at time $-t_{\rm c}$.  
From this time onward to time $t$$=$$0$ they 
will remain coalesced, since the same random
numbers propagate them forward to $t$$=$$0$.  This coalescence 
gives the Propp-Wilson method the `coupling from the past' name.  

Here is the Propp-Wilson prescription for simulation of $N$ Ising spins.  
First obtain a {\it very long\/} history of randomly picked spins 
and random numbers used to compare against ${\rm Prob}(\sigma_i=+1)$.  
Start with all $2^N$ possible spin configurations at time $t$$=$$-1$.  
Propagate all configurations forward in time to $t$$=$$0$.  
Remember, this is done using the {\it same\/} set of random numbers for 
each configuration.  
If {\it all\/} $2^N$ configurations have coalesced 
to the same spin configuration at $t$$=$$0$, 
this is the spin configuration you want.  
If not, start all $2^N$ spin configurations at $t$$=$$-2$, propagate 
forward with the same random number history, 
and see if they have coalesced at $t$$=$$0$.  This is 
continued until at some time $-t_{\rm PW}$ the coalescence of all 
$2^N$ spin configurations has occurred.  The spin configuration 
obtained in this fashion is {\it guaranteed\/} to be an exact 
sampling of the equilibrium spin distribution.  
An exact sample means that the spin configuration is drawn 
according to the probability associated with the equilibrium distribution.  
Note that care must be taken to ensure that the 
long random-number history is never exceeded before the coalescence occurs. 

Why does this give an exact sampling?  
Let us imagine that we have run an infinite ensemble of 
spin configurations from $t$$=$$-\infty$ to $t$$=$$-t_{\rm PW}$ 
using different random numbers for each.  
Since we have run our Markov chain 
infinitely long, we are guaranteed that the spin configurations at 
$t$$=$$-t_{\rm PW}$ are distributed according to the equilibrium 
spin distribution.  Each of the starting spin configurations will 
have evolved into 
one of the $2^N$ spin configurations at $t$$=$$-t_{\rm PW}$.  No matter 
which spin configuration the system is in at time $-t_{\rm PW}$, we know that 
at time $t$$=$$0$ all $2^N$ configurations have coalesced to 
a single spin configuration since we have used our history of random 
numbers which is the same for each configuration to run from 
$t$$=$$-t_{\rm PW}$ to $t$$=$$0$.  
Thus the sample generated is drawn from the equilibrium spin distribution.  

In principle only a coupling-from-the-past method, 
like the Propp-Wilson method, is required to obtain a perfectly 
random sample.  In practice 
another relationship is required since the number of 
states grows very quickly with lattice size.  
If all $2^N$ states could be handled, the partition function could be 
calculated exactly.  
For special 
cases it is possible to introduce a {\bf partial ordering\/} on the 
states.  For the Ising model, we say that two spin configurations ${\vec x}$ 
and ${\vec y}$ have a partial ordering, ${\vec x}\preceq{\vec y}$, if 
for each lattice site the spin in ${\vec x}$ is down whenever 
the spin in ${\vec y}$ is down.  
The transition probability of (\ref{EqProb}) 
preserves the partial ordering 
for ferromagnetic interactions.  
Consequently, 
rather than insist that all $2^{N}$ configurations have 
coalesced, only the two extreme partial ordering states, 
all spins up and all spins down, need 
to be considered.  If these two states have coalesced, then all other 
states have coalesced.  
It is this desired property of partial ordering 
that restricts one to ferromagnetic interactions.  

\begin{figure}[t]
\centerline{
\hspace*{-0.3truecm}
\rotatebox{0}{
\includegraphics[width=0.45\textwidth]{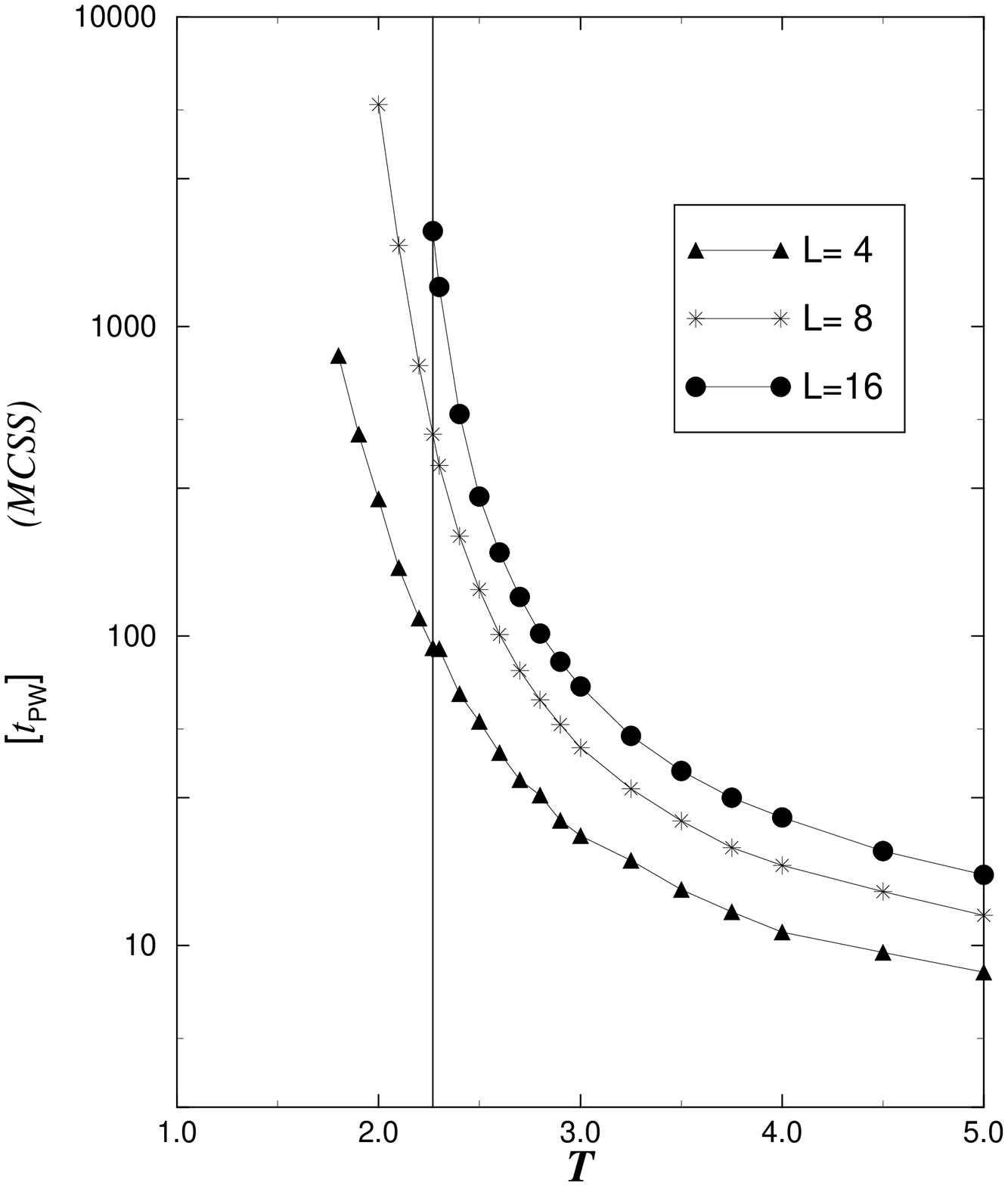}
}
\hspace*{0.4truecm}
\rotatebox{0}{
\includegraphics[width=0.45\textwidth]{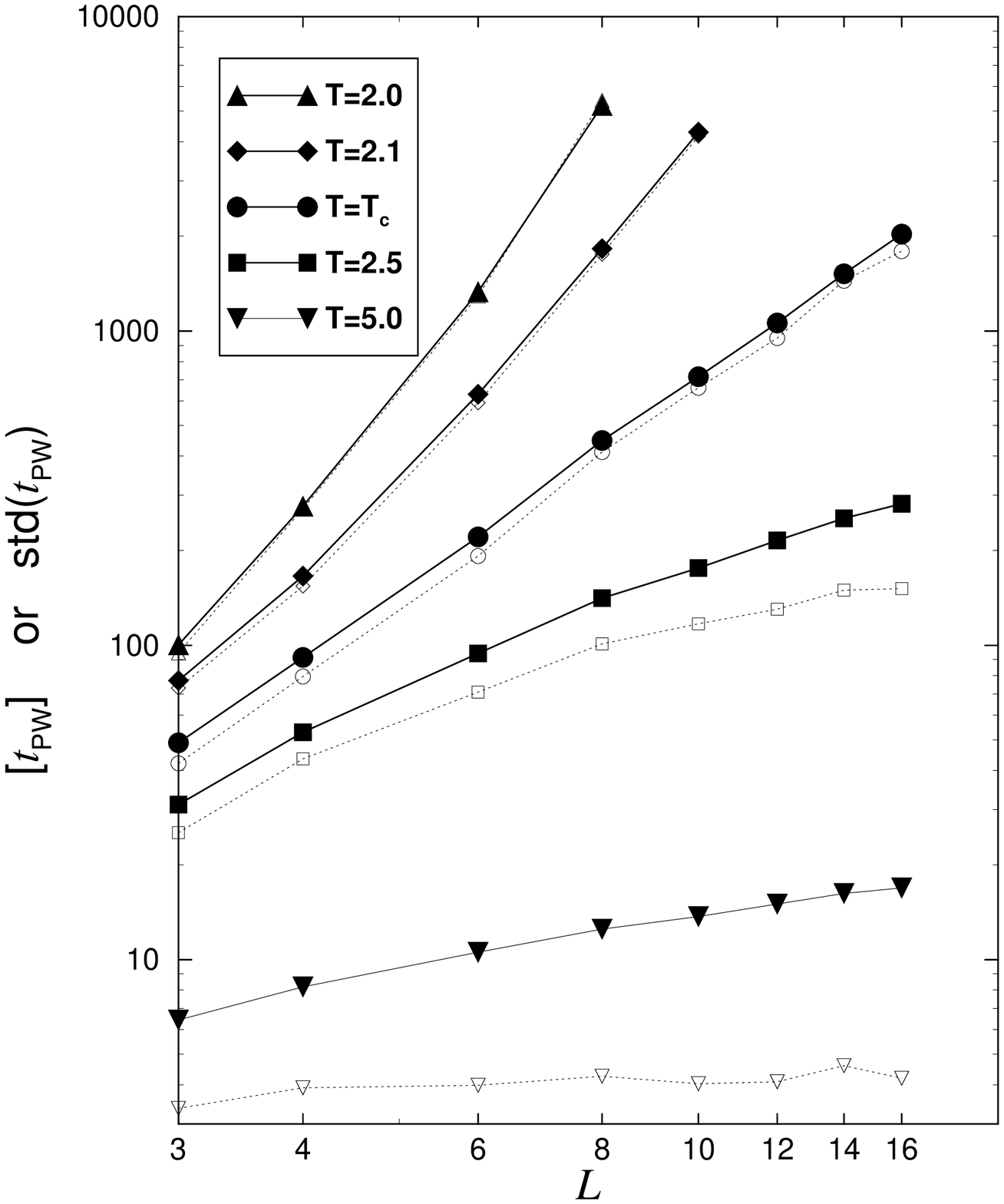}
}
}
\vspace*{0.1truecm}
\caption[]
{
(a) The average Propp-Wilson time, $[t_{\rm PW}]$, is shown as a function of 
temperature for $H$$=$$0$ and $L$$=$$4$, $8$, and $16$.  
The vertical line is the exact value of the critical temperature, 
$T_{\rm c}$.  
The lines are guides for the eye.  
(b) The $L$ dependence on a log-log plot is shown for 
$[t_{\rm PW}]$ (filled symbols joined by solid lines) 
and the standard deviation of $t_{\rm PW}$ (corresponding open symbols 
joined by dashed lines).  Both 
of these grow faster than power laws below $T_{\rm c}$, slower 
than power laws above $T_{\rm c}$, and are expected to grow 
as a power law at $T_{\rm c}$.  
}
\label{fig1}
\end{figure}

The Propp-Wilson method was run to obtain 
$t_{\rm PW}$ as a function of $L$, $T$, and $H$.  
For each value of parameters 1000 exact 
samples were obtained.  
Note that if coalescence occurs at time $-t_{\rm PW}$, it also occurs 
for any earlier time.  This allows a bisection method to be used 
to find $t_{\rm PW}$.  Figure~1(a) shows the average value of 
the Propp-Wilson time, 
$[t_{\rm PW}]$, in Monte Carlo Steps per Spin (MCSS), at $H$$=$$0$.  
Since the phase space grows with 
system size, $[t_{\rm PW}]$ increases with $L$.  Furthermore, 
since it takes longer to sample all of phase space at lower 
temperatures, $[t_{\rm PW}]$ increases as $T$ is lowered.  

For a $d$ dimensional system with correlation length $\xi$, 
below but near $T_{\rm c}$ the ergodic time is expected to scale 
as $\tau_{\rm e}$$\propto$$L^z \exp\left[{\rm const} (L/\xi)^{d-1}\right]$ 
\cite{refBH}.  
Here $z$ is the exponent associated with critical slowing down.  
For $T$$\ge$$T_{\rm c}$ the `intrinsic' order parameter relaxation 
time $\tau$ is the longest relaxation time in the system \cite{refBH}, so 
above $T_{\rm c}$ one expects 
$\tau$$\propto$$\xi^z$$\propto$$|1-T/T_{\rm c}|^{-\nu z}$ where 
$\nu$ is the correlation length exponent.  Right at $T_{\rm c}$ this 
yields $\tau$$\propto$$L^z$.  
Figure~1(b) shows the $L$ dependence of $[t_{\rm PW}]$ 
for $H$$=$$0$ and various temperatures.  The graph illustrates that 
this time grows slower (faster) than a power law above (below) $T_{\rm c}$. 
At $T_{\rm c}$ this time seems to scale as a power law.  However, care 
must be used in associating $[t_{\rm PW}]$ 
with either $\tau$ or $\tau_{\rm e}$.  
One might expect that $[t_{\rm PW}]$ yields an upper bound on 
the larger of $\tau_{\rm e}$ or $\tau$.  
As seen in Fig.~1 
even at high temperatures $[t_{\rm PW}]$ 
increases with $L$.  These additional size 
dependent factors in $[t_{\rm PW}]$ 
might also be expected to be present 
at lower temperatures.  
In a recent thesis \cite{refFis} $[t_{\rm PW}]$$\propto$$L^\Omega$ was 
postulated, and the value $\Omega$$\approx$$2.44$ was found with 
system sizes $L$ up to $310$.  (This assumes that the 
presented values are at $T_{\rm c}$, which was not stated explicitly.)  
This is somewhat high compared to the accepted value of $z$$\approx$$2.21$ 
\cite{refzzz}.  It is also possible to calculate higher 
moments of $t_{\rm PW}$.  Figure~1(b) shows the standard deviation 
per spin for $t_{\rm PW}$ as a function of $L$.  
The behavior seems to be close to 
those of $[t_{\rm PW}]$.  Note that the standard deviation becomes 
very large, which necessitates using a large number of samples to obtain 
a reasonable value for $[t_{\rm PW}]$.  

\begin{figure}[t]
\centerline{
\hspace*{-0.3truecm}
\rotatebox{0}{
\includegraphics[width=0.45\textwidth]{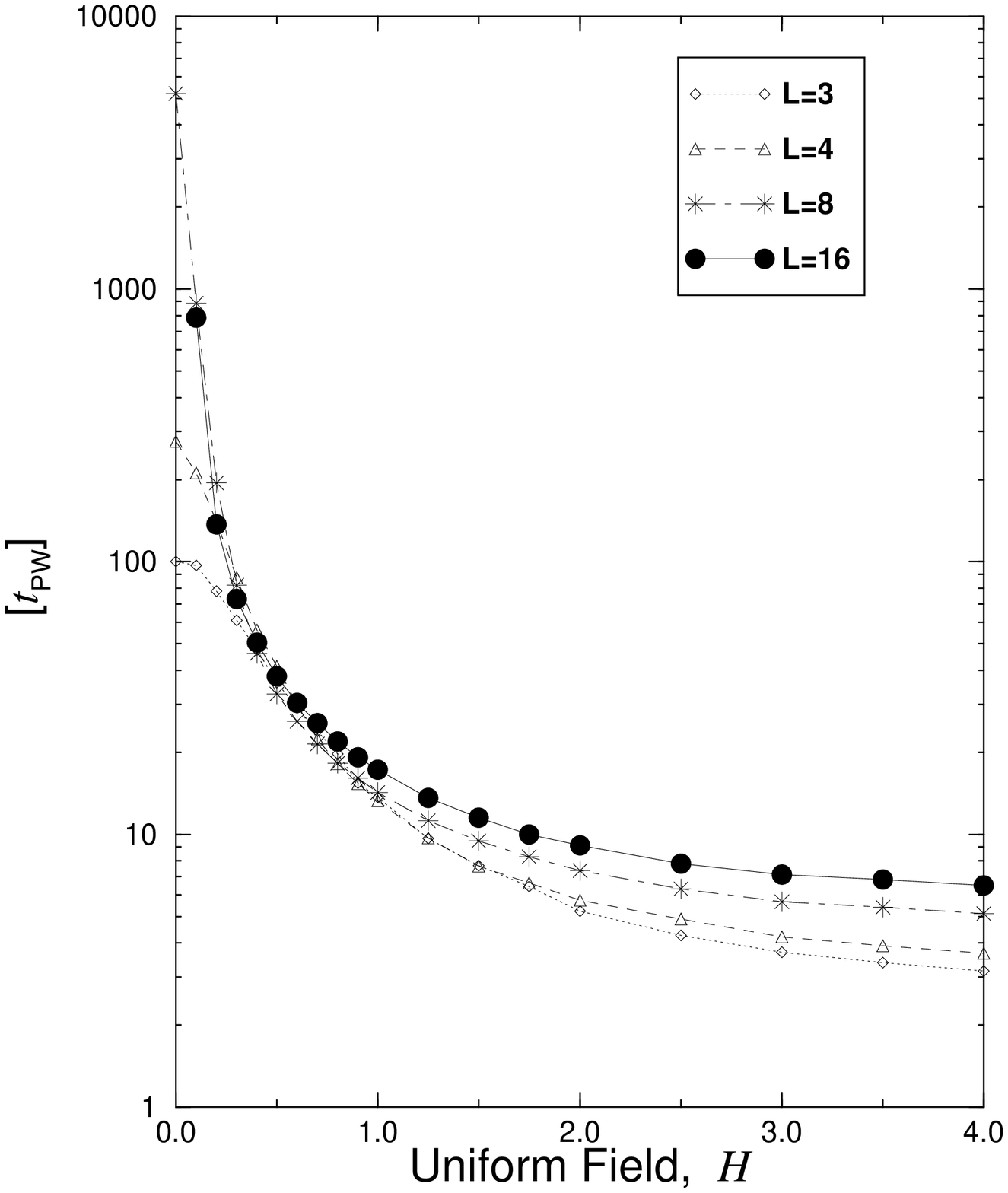}
}
\hspace*{ 0.4truecm}
\rotatebox{0}{
\includegraphics[width=0.45\textwidth]{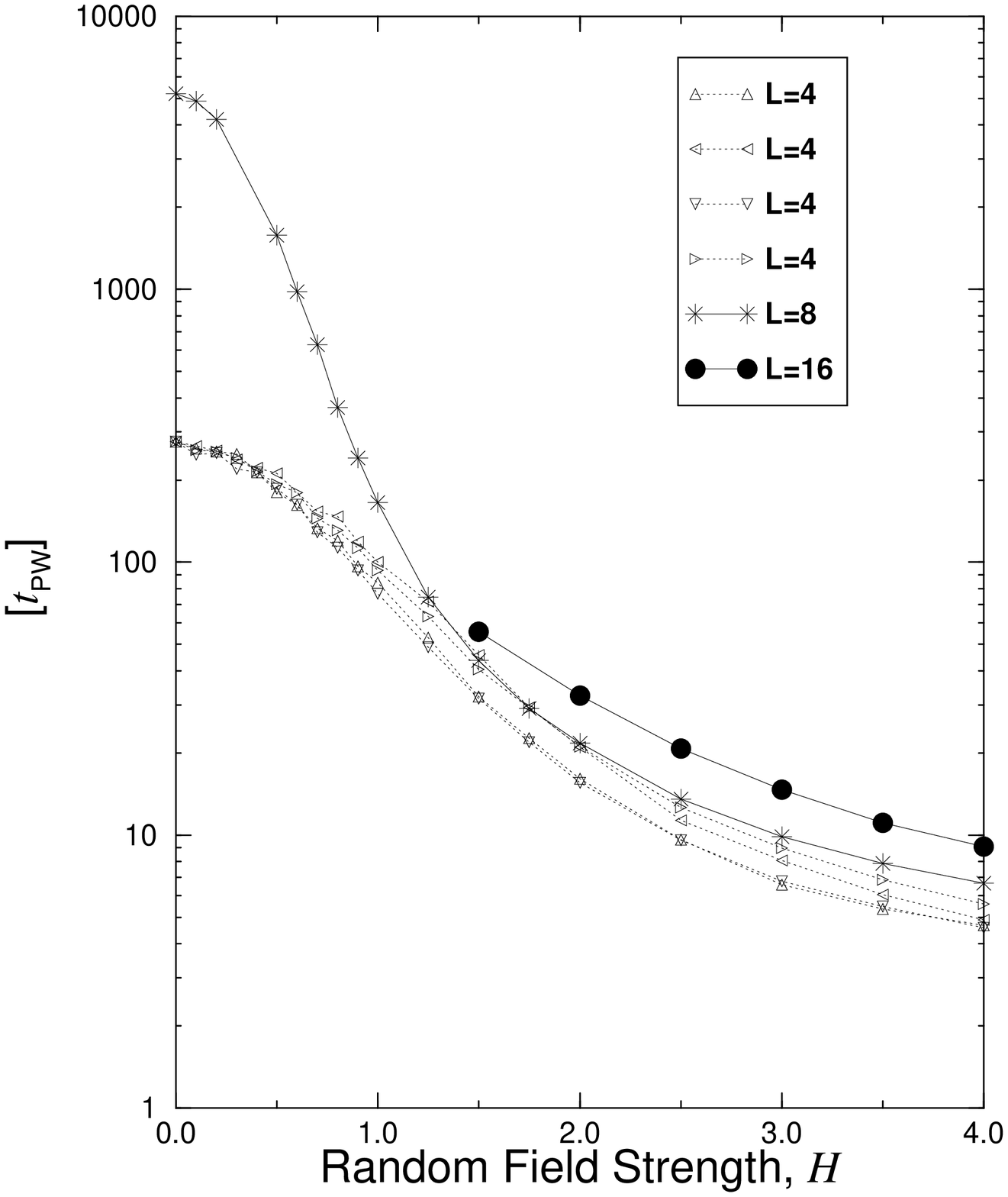}
}
}
\vspace*{1.0truecm}
\caption[]
{
The average Propp-Wilson time, $[t_{\rm PW}]$, in 
MCSS is shown at $T$$=$$2.0$ as a function of 
$H$.  
The lines are guides for the eye.  
Note that the points at $H$$=$$0$ are the same for both plots.  
(a) Is for a 
uniform magnetic field.  For $L$$=$$16$, $H$$=$$0.1$ 
is the smallest value shown.  
(b) Is for quenched randomly distributed 
fields equal to $-H$ or $+H$.  
Four different distributions of fields are shown for $L$$=$$4$, 
and one distribution for $L$$=$$8$ and $L$$=$$16$.  For 
$L$$=$$16$, $H$$=$$1.5$ is the smallest value shown.  
}
\label{fig2}
\end{figure}
Figure~2(a) shows $[t_{\rm PW}]$ 
at $T$$=$$2.0$ for different values of 
$L$ and uniform applied field, $H$.  At strong 
fields the larger system sizes have larger values of $[t_{\rm PW}]$, 
for intermediate fields all system sizes have similar values of 
$[t_{\rm PW}]$, and for weak fields the smaller system sizes 
break away from this curve and go to their maximum 
at $H$$=$$0$.  This behavior is reminiscent of the cross-overs 
encountered in the study of metastability, where for $T$$<$$T_{\rm c}$ 
as the field is decreased different regimes 
are encountered in a finite system \cite{RNKR}.  

To illustrate the application of the Propp-Wilson method to a 
random-field Ising model, Fig.~2(b) shows data for $[t_{\rm PW}]$ 
as a function of the strength of the random field 
for various quenched field configurations for $L$$=$$4$, and one 
quenched field configuration for $L$$=$$8$ and $16$.  The trends 
seem to be similar to those of Fig.~2(a).  However the values 
for $[t_{\rm PW}]$ seem to increase faster as $H$ decreases than 
in the uniform field case.  This is in agreement with the 
physical idea that a system with random quenched disorder 
has a longer ergodicity time than one without disorder.  
For more discussion of random fields see references 
\cite{refEx,refPM,refnond}.  

In summary, it is possible to utilize coupling-from-the-past ideas 
to obtain perfect samples of Ising spin configurations from the 
equilibrium spin distribution.  
To use the ideas in practice, however, another relationship is essential,  
that of a partial ordering preserved by the transition probabilities.  
This restriction 
allows one to study random fields, but 
no partial ordering is known for systems with antiferromagnetic bonds.  
The value for the average and standard deviation for 
$t_{\rm PW}$ seem to have reasonable dependences on 
system size, field strength, and temperature if $t_{\rm PW}$ is 
viewed as an upper bound for the ergodic time of the system.  

It is possible to use cluster dynamics instead of 
single-spin-flip dynamics.  Since this decreases the ergodic time 
near $T_{\rm c}$, this allows one to obtain perfect samples for 
large systems --- so far the largest single example I have 
seen had $L$$=$$2100$ in a preprint by Propp and Wilson.  
It has also been shown that the 
expected running time for the Propp-Wilson method 
should be within a constant multiple of 
the maximum expected time for two states to coalesce in the 
Markov chain, and that this running time is near optimal 
for exact sampling.  Thus the ability to obtain results for 
large systems at temperatures above and near $T_{\rm c}$ 
for models that have a partial ordering seems realistic.  
Unfortunately, these partial orderings are extremely difficult 
to find in more general models.  Also, the long expected run 
times seem to make the application of these types of algorithms 
rather limited in studies of low-temperature behavior for 
systems with quenched randomness and high barriers.  

%
This work is supported in part by the NSF through grant number 9871455, 
and through the Supercomputer Computations Research Institute which is 
funded by the U.S.\ DoE and the State of Florida.  

\bibliography{/a/alpha2/home/scri42/users/rikvold/%
decstation/tex/biblio/metastab}
\bibliographystyle{unsrt}


\clearpage
\addcontentsline{toc}{section}{Index}
\flushbottom
\printindex

\end{document}